\documentclass[a4paper,10pt,aps,prb,english,twocolumn,bibnotes,altaffilletter]{revtex4}
\usepackage{amssymb}
\usepackage{amsmath}
\usepackage{babel}
\usepackage{graphicx}
\usepackage{natbib}
\usepackage{bm}
\usepackage{epsfig}

\newcommand{\be}{\begin{equation}}
\newcommand{\ee}{\end{equation}}

\newcommand{\opv}{{\bf Op}}
\begin{document}
\title{A simple derivation of the Lorentz transformation and of the related velocity and 
acceleration formulae}
\author{J.-M. L\'evy}  
\email{jmlevy@in2p3.fr}
\affiliation{Laboratoire de Physique Nucl\'eaire et de Hautes Energies,
CNRS - IN2P3 - Universit\'es Paris VI et Paris VII, Paris.}
\begin{abstract}
The Lorentz transformation is derived from the simplest thought experiment by using the simplest vector 
formula from elementary geometry. The result is further used to obtain general velocity and acceleration 
transformation equations.
\end{abstract}

\maketitle
\section{Introduction}
Many introductory courses on special relativity (SR) use thought experiments in order to 
demonstrate time dilation and length contraction from Einstein's two postulates by using 
conceptual devices like the well known light clock or variants thereof (see below or 
e.g.\cite{Krane} An extensive bibliography is given in \ \cite{Mathews})
However, once these two effects are established, most authors return to the postulates to 
derive the Lorentz transformation (LT), taking the route which is usual in advanced texts but which is certainly not the easiest one 
to begin with. \\
However, deriving the LT directly from these effects is possible and has obvious advantages 
for beginners. It allows, for example, to bypass the use of group structure and 
linearity. Important as though they are in fundamental physics, dispensing with these 
considerations allows for a very direct first contact with the conceptually demanding subject of 
SR.\cite{Levy} \ More elaborate derivations from fundamental principles can be left for a 
second pass.\\

In the present article, we show that the LT can be derived from length contraction through a purely geometrical 
argument which amounts to expressing the basic vector addition formula in two frames 
in rectilinear and uniform relative motion.
 This reasoning leads to a very simple and possibly new way of writing the space part of the LT, 
which in turn allows for an easy derivation of the velocity and acceleration tranformations.\\
This type of derivation was used already in a paper published in this Journal a long time ago.
\cite{Park}\  However, the author of this paper missed what we think is the easiest way to derive 
the time transformation formula and obtained it through a rather contrived argument, 
introducing an artificial extension of the 'time interval'. Also, as in most papers on the subject,
the derivation was limited to transformations 
between two reference frames in the so-called 'standard configuration', \cite{Rindler} viz. parallel axes, $OX'$ 
sliding along $OX$ with co\"{\i}ncident space-time origins.\\

The present paper is organised as follows: in order to prevent possible objections which are often not taken care of 
in the derivation of the two basic effects using the light clock, we start by reviewing it briefly in 
section II. The LT between two frames in 'standard configuration' is first derived from length 
contraction in 
section III. Section IV treats the more general case of an arbitrarily oriented relative 
velocity. In section V we use the expression obtained in section IV to find the velocity 
and acceleration transformation. Section VI contains 
our summary and conclusions.\\  

\section{Time dilation and length contraction}
\subsection{The light clock}
The light clock is the conceptual device sketched on Fig.1 : a light signal bounces back and forth between two
parallel mirrors maintained a constant distance apart with the aid of pegs (not drawn). The signal triggers the 
registering of a tick each time it hits the 'lower' mirror (Fig.1 left). We thus have a perfect clock with period 
\be T_0 = \frac{2L_0}{c} \ee with $L_0$ the distance between the mirrors and $c$ the speed of light. \\

\begin{center}
Fig.1 The light clock at rest (left) and moving(right)
\end{center}
\subsection{Time dilation}
Let's now look at the clock in a frame wherein it travels at a constant speed $v$ in a direction parallel to the 
the mirrors. We might assume that the mirrors are constrained to slide in two parallel
straight grooves which have been engraved a constant distance $L_0$ apart, so that there
cannot be any argument about a variation of the pegs length when they are moving.\label{TD}\\ 
By the first postulate, 
this moving clock must have the same period in its rest frame than its twin at rest in the laboratory.\\
On the other hand, the length traveled by the signal in the lab is longer than the 
length it travels in the clock rest frame (see Fig.1 right) If $T$ is the interval between two ticks in the lab, then 
by Einstein's second postulate and the Pythagorean theorem, we have that \be (c T/2)^2 = L_0^2 
+ (v T/2)^2 \ee from which 
\be T = \frac{T_0}{\sqrt{1-(\frac{v}{c})^2}} = \gamma T_0 \ee follows, showing that the moving 
clock runs more slowly in the lab than its stationnary twin. The second equal sign defines the ubiquitous {\it Lorentz $\gamma$ 
factor}.
\subsection{Length contraction} Now the moving clock is traveling in a direction perpendicular to 
the plane of its mirrors relative to the lab observer. In this case, no check can be kept of the 
inter-mirror distance. To make sure that (for the same $v$) the clock period hasn't changed, we 
can imagine it accompanied by an identical second clock oriented as before with respect to its lab 
velocity. 
Both clocks have the same period in their common rest frame and by the argument already 
given, the clock moving parallel to its mirrors has period 
$T$ in the lab (cf. above); therefore we can be sure that the clock moving 
perpendicularly to its mirrors also has period $T$ in the lab frame.
Anticipating the result which will be forced upon us, let $L$ be the inter-mirror distance as 
measured in the laboratory frame.
Now consider the time taken by the light signal to make its two-way travel in the laboratory 
frame; starting from the rear mirror (which was the 'lower' mirror before the clock was 
rotated), it will reach the front 
mirror after a time $t$ given by $ct = L +vt$ and will need a further time 
lapse $t'$ given by $vt' = L-ct'$
for the return leg, which makes a total of $\frac{L}{c(1-\frac{v^2}{c^2})}$
Equating this expression with the one already obtained for $T$, one is forced to conclude 
that \be L = L_0\sqrt{1-(\frac{v}{c})^2} = \frac{L_0}{\gamma}\ee
 \\
That the distances in the directions orthogonal to the motion are not changed can be demonstrated 
by invoking grooves arguments like the one we used for the time dilation derivation. For 
example, we can imagine that the rims of the mirrors are fitted with skates gliding
perpendicularly to the mirror planes in two parallel straight grooves.  
\section{Lorentz transformation along the $x$ axis}
Let us now envision two frames in 'standard configuration' with $K'$ having velocity $\bf v$ with respect to $K$ and 
let $x,t$ (resp. $x',t'$) be the coordinates of event $M$ in the two frames. Let $O$ and $O'$ be the spatial origins 
of the frames; $O$ and $O'$  co\"{\i}ncide at time $t = t' = 0$\\

Here comes the pretty argument: all we have to do is to express the relation \be {\bf OM = OO' + O'M} \ee between 
vectors (which here reduce to oriented segments) in both frames.\\

In $K$, $\overline{OM} = x$, $\overline{OO'} = vt$ and $\overline{O'M}$ {\it seen from $K$} is $\frac{x'}{\gamma}$ 
since $x'$ is $\overline{O'M}$ as measured in $K'$
Hence a first relation: \be x = vt + \frac{x'}{\gamma} \label{eq1} \ee 
In $K'$, $\overline{OM} = \frac{x}{\gamma}$ since $x$ is $\overline{OM}$ {\it as measured in $K$}, $\overline{OO'} = 
vt'$ 
and $\overline{O'M} = x'$. Hence a second relation: \be \frac{x}{\gamma} = vt' + x' \label{eq2} \ee 
Relation (\ref{eq1}) yields immediately \be x' = \gamma(x-vt) \label{tr1} \ee which is the 
$x$-axis 'space' part of the LT 
and relation (\ref{eq2}) yields the inverse \be x = \gamma(x'+vt') \label{tr2} \ee of this 'space part'. Eliminating 
$x'$ between (\ref{tr1}) and (\ref{tr2}) quickly leads to the formula for the transformed time: \be t' = 
\gamma(t-vx/c^2) \ee the inverse of which 
could easily be found by a similar elimination of $x$.\\
Coordinates on the $y$ and $z$ axes are unchanged for the already stated reason that distances 
do not vary in the directions perpendicular to the velocity. 
\section{The case of an arbitrary velocity}
In the following, $\bf v$ will denote the velocity vector of $K'$ w.r.t. $K$ and
$\bf r$ (resp. $\bf r'$) the position vector of the event under consideration as measured in frame $K$ (resp $K'$). 
We further define \be \bf u = \frac{v}{|v|} \ee the unit vector parallel to $\bf v$. \\
From our findings of section 2, we see that only the component of $\bf r$ parallel to $\bf v$ is affected when 
looking at it from the other frame, while the normal components are unchanged. We resolve $\bf 
r$ into parallel and 
perpendicular components according to \be \bf r = u u.r +(1 - u\otimes u)r = r_{\parallel} + r_{\perp} \ee where the 
dot stands for the 3-space scalar product, $\bf 1$ is the identity operator and $\bf u\otimes u$ is the dyadic which 
projects out the component parallel to $\bf u$ from the vector it operates upon, viz \be \bf (u\otimes u)V = (u.V) 
u \ee  
\\ The operator which 
contracts the projection on $\bf u$ by $\gamma$ while leaving the orthogonal components 
unchanged must yield: \be 
\bf u \frac{u.r}{\gamma} + (1 -u\otimes u)r = (1 +\frac{{\rm 1}-\gamma}{\gamma}u\otimes u)r \ee  Let us therefore 
define \be \bf Op(\gamma^{\rm -1}) = 1 + \frac{{\rm 1}-\gamma}{\gamma} u \otimes u \label{gm1} \ee 
The inverse operator  must correspond to multiplication of the longitudinal part by $\gamma$ and is therefore
\be \bf Op(\gamma) = Op(\gamma^{\rm -1})^{\rm -1} = 1 + (\gamma-{\rm 1}) u\otimes u \label{g} \ee as can also be 
checked by multiplication of the right-hand sides of (\ref{gm1}) and (\ref{g}) . Note that these operators are even 
in $\bf u$ and therefore independent of the orientation of $\bf v$.\\

Mimicking what has been done in section 3, let us 
write again \be \bf OM = OO' + O'M \ee but for vectors now, taking 
care of the invariance of the orthogonal parts. We get in frame $K$: \be \bf r = v{\rm t} + Op(\gamma^{\rm -1})r' 
\label{K} \ee and in frame $K'$: \be \bf Op(\gamma^{\rm -1})r = v{\rm t'} + r' \label{K'} \ee
 
Using (\ref{g}) relation (\ref{K}) yields immediately: 
\be \bf r' = Op(\gamma)(r - v{\rm t}) = (1 + (\gamma-{\rm 1}) u\otimes u)(r - v{\rm t}) \label{Lor1} \ee which is 
probably the simplest way to write the space part of the rotation free homogenous LT. The usual $\gamma$ factor of the 
one dimensionnal transformation is simply replaced by the operator $\bf Op(\gamma)$\\
By substituting (\ref{Lor1}) into (\ref{K'}), we find:
\be \bf Op(\gamma^{\rm -1})r = v{\rm t'} + Op(\gamma)(r-v{\rm t}) \ee or, using \be \bf Op(\gamma)v = \gamma v \ee 
and with the explicit form of $\bf Op$:
\be (\frac{1-\gamma}{\gamma} -(\gamma-1))\frac{\bf v v.r}{v^2} +\gamma {\bf v} {\rm t} = {\bf v}{\rm t'} 
\label{mess} \ee  
Using now 
\be 1-\gamma^2 
= 
-(\frac{v}{c})^2\gamma^2 \ee and crossing away $\bf v$ on both sides, (\ref{mess}) yields:
\be t' = \gamma (t-\frac{\bf v.r}{c^2}) \label{Lor2} \ee i.e. the time transformation equation. 
\section{Velocity and acceleration transformations}

\subsection{Velocity}
The two formulas thus obtained for the L.T. are so simple that they can readily be used to yield the
velocity transformation equation without the need of complicated thought experiments and algebraic manipulations.
Differentiating (\ref{Lor1}) and (\ref{Lor2}) w.r.t. $t$ and taking the ratio of the 
equalities thus obtained yields,
 \begin{eqnarray} ( {\rm with \;} {\bf V'} = \frac{d{\bf r'}}{dt'} & {\rm and} & {\bf V} = \frac{d{\bf r}}{dt})     
\end{eqnarray} 
 \be \bf V' = \frac{\rm 
1}{\gamma}\frac{({\rm 1} + (\gamma-{\rm 1}) u\otimes u)(V - v)}{{\rm 1}-\frac{v.V}{\rm c^2}} 
\label{Lor3} \ee which 
is the general velocity transformation formula. 
\subsection{Acceleration}
The compact $\bf Op$ notation helps to keep the algebra tidy when differentiating (\ref{Lor3}) 
w.r.t. $t$; dividing the derivative of (\ref{Lor3}) by that of (\ref{Lor2}) one 
finds \be \bf A' = \frac{\rm 1}{\gamma^2}\frac{Op(\gamma) A ({\rm 1}-\frac{v.V}{\rm c^2}) 
+Op(\gamma)(V-v)\frac{v.A}{\rm c^2}}{({\rm 1}-\frac{V.v}{\rm c^2})^{\rm 3}} \ee
Expliciting $\bf Op$, simplifying and regrouping terms, one obtains after a page of algebra: 
\be \bf A' 
= \frac{A-\frac{\gamma}{\gamma+{\rm 1}}\frac{v.A v}{\rm c^2}+\frac{v \times (V \times A)
}{\rm c^2} }{\gamma^{\rm 2} ({\rm 1}-\frac{V.v}{\rm c^2})^{\rm 3}} \label{Lor4} \ee
By making the necessary substitutions: $\bf V \rightarrow u'$, $\bf V' \rightarrow u$, $\bf v \rightarrow -V$ and 
specializing to $\bf V$ parallel to $Ox$, one can easily check that the component equations derived from (\ref{Lor4})  
agree with those published in.\cite{Mathews}\\
As an example of use of this acceleration transformation, we take $\bf V=v$ and $\bf v.A = 
0$, and obtain $\bf A' = \gamma^{\rm 2} A$ retrieving the known result that a particle in a 
circular storage ring undergoes a proper ($\bf A'$) acceleration that is a factor $\gamma^2$ 
larger than the lab ($\bf A$) acceleration.(\cite{Rindler3}) Moreover, the 
two accelerations are parallel, which is far from obvious a priori. Observe that all the terms which 
can make $\bf A'$ and $\bf A$ different in direction as well as in magnitude vanish in the $c 
\rightarrow \infty$ limit, consistent with the fact that acceleration is an invariant quantity 
under a change of inertial frame in newtonian physics.  \\
Setting $\bf V = v$ and taking $\bf v$ parallel to $\bf A$ we also retrieve another known fact: a particle in
rectilinear motion undergoes a proper acceleration which is larger than its lab acceleration 
by a factor $\gamma^3$. \\[.2cm]
These two examples are but special cases of a general formula connecting proper acceleration
and acceleration in the laboratory frame, which can be obtained by setting $v = V$ in
(\ref{Lor4}), viz. \be {\bf A'} = \gamma^2{\bf Op}(\gamma)\bf A \label{Lor5} \ee
Here $\gamma$ and $\opv(\gamma)$ are calculated using the laboratory velocity of the
accelerating body that is also the velocity of the inertial frame in which it is 
instantaneously at rest. Equation (\ref{Lor5}) can be readily inverted to yield the laboratory 
acceleration given the proper acceleration, if needed.

\section{Summary and conclusion}
We have shown that the general rotation free homogenous LT can be derived once length contraction 
has been established by writing the elementary vector relation (sometimes dubbed 'Chasles' relation) ${\bf OM = OO' + 
O'M}$ in the two frames considered.\cite{MacDo}  The extension from the special one dimensional case to the 
3-dimensional case is completely straightforward. 
The relation we have obtained allows for a simple derivation of the velocity and 
acceleration transformations without the need for complicated thought experiments and 
algebraic manipulations.
\begin{footnotesize}

\end{footnotesize}


\begin{thebibliography}{9}
\bibitem{Levy} For the sake of completeness, it must be mentionned that the LT can in fact be 
derived by using the sole relativity principle (first Einsein postulate) and dispensing with the 
second (invariance of the speed of light) This is certainly most appealling from a puristical 
viewpoint, but it demands more abstract 
work which is what we are trying to avoid here, having in mind beginning students.
\end{thebibliography}
\end{document}